\documentclass{article}
\usepackage{style/stylo}
\usepackage{graphicx} 

\title{War Elephants: Rethinking Combat AI and Human Oversight}
\author {
	{\small{
			Philip Feldman\textsuperscript{\rm *}
			Aaron Dant \textsuperscript{\rm *}
			Harry Dreany \textsuperscript{\rm **}
	}}
}
\date{\today}

\begin{document}
	
	\maketitle
	\raggedright
	\textsuperscript{\rm *} {\small{ASRC Federal, Beltsville, Maryland, USA}} \\
	\textsuperscript{\rm **} {\small{Marine Corps Warfighting Lab, Quantico, Virginia USA}}
	
	\begin{abstract}
    \normalsize
    \noindent
    This paper explores the changes that pervasive AI is having on the nature of combat. We look beyond the substitution of AI for experts to an approach where complementary human and machine abilities are blended. Using historical and modern examples, we show how autonomous weapons systems can be effectively managed by teams of human \textit{AI Operators} combined with AI/ML \textit{Proxy Operators}. By basing our approach on the principles of complementation~\cite{krakowski2022artificial}, we provide for a flexible and dynamic approach to managing lethal autonomous systems. We conclude by presenting a path to achieving an integrated vision of machine-speed combat where the battlefield AI is operated by AI Operators that watch for patterns of behavior within battlefield to assess the performance of lethal autonomous systems. This approach enables the development of combat systems that are likely to be more ethical, operate at machine speed, and are capable of responding to a broader range of dynamic battlefield conditions than any purely autonomous AI system could support.



\end{abstract}
	\section{INTRODUCTION}
\label{sec:introduction}

In November of 2022, the unveiling of ChatGPT marked a pivotal moment in the unfolding story of artificial intelligence (AI). For decades, AI had been a subject of intense study, but for the average person, it remained an abstract concept, present more science fiction or technology demonstrations like IBM's Watson winning Jeopardy. However, with ChatGPT, artificial intelligence stepped out of the lab and people could engage in conversations with it in the same way they would interact with their friends and coworkers. ChatGPT shifted the public perception of AI from an future possibility to a tangible reality.

Simultaneously, thousands of miles away from the offices where ChatGPT was developed, Ukraine was integrating artificial intelligence into the theater of war. In the face of massive Russian RF jamming, Ukraine deployed AI-enhanced drones capable of operating effectively without human supervision under hostile conditions. This application of AI was not about conversing or simulating human-like interactions; it was about enhancing operational capabilities against larger adversary where traditional systems faltered. The contrast between ChatGPT's cognitive prowess and the utilitarian application of AI in Ukraine's drones illustrate the versatile and transformative nature of artificial intelligence across vastly different domains.

These two applications of AI have led to a potential shift in the strategic thinking of the Department of Defense (DoD) regarding battlefield AI. Initially AI doctrine emphasized the paramount importance of \enquote{appropriate levels of human judgement} over AI systems, with \enquote{clear procedures to activate and deactivate system functions, and to provide transparent feedback on system status}~\cite{dod:3000-09_2023}. Requirements for \enquote{human-machine interface for autonomous and semi-autonomous weapon systems}~\cite{dod:3000-09_2023} imply a level of direct human interaction, often known as human-in-the-loop (HITL) or human-on-the-loop (HOTL). This direct involvement of humans either participating directly with the AI or monitoring for errors is a testament to the cautious approach towards automating critical processes. 

This shift is happening because of the realization that AI, with its inevitable technological ubiquity, will not merely supplement but fundamentally alter operations and intelligence gathering on the battlefield as it already has started to almost everywhere else in society. This is not the first time such an evolution has happened. Just in the last generation, the advent of the internet and the proliferation of smartphones revolutionized communication and access to information the the extent that the DoD has had no choice but to adapt.  Much like these innovations, AI is on a trajectory to embed itself at a similarly fundamental level.

The challenge now lies not in trying to dictate the integration of AI but in adapting to its inevitable pervasiveness, ensuring the reliability and effectiveness of military systems amidst concerns of data biases, computational \enquote{hallucinations} while ensuring that LAWS behaves in accordance with International Humanitarian Law (IHL). This document aims to explore these challenges and propose pathways to mitigate them, recognizing that the journey towards AI-augmented warfare is not only inevitable but already underway.
	\section{BACKGROUND}
\label{sec:background}

Throughout history, humans have leveraged the innate capabilities of autonomous and semi-autonomous systems in peace and war. Lessons from this period of partnerships with animals may provide the greatest insights into trustworthy interactions with AI and LAWS. We are now creating a world of omnipresent of artificially intelligent systems. The military must inevitably reflect this transition in greater society. 

There is a rich history of sophisticated interactions with animals; horses as self-driving vehicles, dogs as remote manipulators, pigeons as communications networks, and elephants as lethal autonomous weapons systems. In each of these cases, there was an understanding that the handlers of these animals, with their expert understanding, had appropriate control of their charges. For example, under the principle of \textit{scienter}, if an animal is known to behave dangerously and causes injury, the owner or handler is legally responsible~\cite{chevalier2007civil}. The understanding that appropriate control can be exerted over animals, even dangerous ones, has been understood for centuries.

Those responsible for the care and management of animals employed for specific tasks have been recognized by titles that reflect their expertise. Falconers train and hunt with birds of prey. Teamsters originally handled draft animals, crucial for transportation purposes. Fanciers breed and train homing pigeons, once essential for long-distance communication. Wranglers are expert in  the care and management of horses~\cite{salter2018animals}. Lastly, mahouts create a unique bond for control over their elephants~\cite{ranking1826historical}. These specialized roles continue to hold significance in both civilian and military applications of working animals today.

Within the annals of military history, dogs and horses have been frequently utilized in warfare. For example, in 7BC, the citizens of of Magnesia, an ancient Greek city-state, employed of animal and human warriors: each combat unit comprised a horse-mounted warrior,  a spear bearer, and a war dog. First, dogs would charge and disrupt the enemy lines. Once in range, this was followed by spear volleys. Last was a cavalry charge, leveraging the full strength of that era's combat system~\cite{baker2022should}.

In the Middle Ages, horses were a critical part of the weapons system that was the armored knight.  War horses were taught specialized behaviors such as biting, kicking, and trampling adversaries. This training ensured that the horses could hold their own in a melee, augmenting the knight's lethality~\cite{gassmann2018combat}.

Interacting with AI systems shares similarities with training and working with animals; both processes are filled with complexities. For instance, just as it requires effort and skill to train a horse to act reliably in a combat situation, teaching an AI to consistently identify the right target or generate accurate text without making mistakes presents its unique set of challenges. One common error in text generation by AI is known as \enquote{hallucinations,} where the AI confidently generates credible but false information~\cite{feldman2023trapping}. Similarly, when AI is used to create images, it often makes mistakes such as depicting people with an incorrect number of fingers. The root of these issues lies in the nature of deep learning systems, which operate on a principle of stochasticity or rule-based randomness~\cite{frankle2018lottery}. They excel at patterns, but often struggle with unique specifics. This inherent feature, while enabling AI to make generalizations, also inevitably leads to errors.

The inherent complexities in AI systems are a product of processes much like the natural variations that make even siblings from the same litter of dogs unique from one another. These systems begin with randomness and employ statistical learning, leading to a level of unpredictability in their operations that can be reduced but not eliminated. Consequently, in sectors crucial to civilian safety and defense, there arises a significant need for specialists or \enquote{AI Operators.} Like the animal handlers before them, these Operators would possess the expertise to understand the behaviors of AI systems and direct then to deliver optimal outcomes. 

The Operator's skill could ensure that the AI's output is reliable and can be trusted by those who rely heavily on such technology, such as the warfighter in the field. Reflecting the growing recognition of \enquote{AI handlers}, the civilian job market is beginning to see the emergence of roles specifically designed for these tasks, with titles such as \enquote{AI Whisperers} and \enquote{Prompt Engineers} signaling the rise of professionals dedicated to managing and mitigating the intricacies of AI systems~\cite{mesko2023prompt}.

The partnership between operator and AI is not an entirely new concept but rather an evolution of historical relationships between humans and intelligent systems, whether organic or artificial. The emergence of specialized roles to facilitate this partnership parallels the ancient practices of training and managing war elephants, underscoring a timeless human endeavor to master sophisticated tools for strategic advantage. This historical context bridges our understanding from the modern digital era back to the times when flesh and blood, rather than circuits and code, formed the foundation of our most advanced technologies, highlighting a continuum in the quest for enhanced operational effectiveness through expert guidance and control.

Next we turn our attention to one of the most formidable human-animal partnerships in history: the war elephants and their mahouts. This ancient association offers profound lessons for contemporary AI use, particularly in understanding the dynamics of control, training, and collaboration between man and beast, principles that resonate with the challenges of guiding AI systems today. The bond between a war elephant and its mahout was built on deep understanding, mutual respect, and meticulous training. By examining this historic example, we will explore how the nuances of this relationship can inform and enhance our approach to developing, managing, and deploying AI technologies, especially in areas where precision, reliability, and adaptability are paramount.

\section{LESSONS FROM HISTORY: WAR ELEPHANTS}



The deployment of elephants in warfare offers a compelling insight into the early use of autonomous systems on the battlefield\footnotemark. These majestic creatures share several characteristics with modern combat AI, highlighting the perennial challenge of integrating complex agents into military operations. Both elephants and AI systems require rigorous training to perform their designated roles effectively. However, their performance is subject to biases -- elephants were influenced by their handlers' methods and the environment they were accustomed to, while AI's biases stem from the data it is trained on and the environment it is deployed in, emphasizing the need for careful preparation and oversight~\cite{fjelland2020general}.

\footnotetext{The examples used in this section are based on the book \textit{War Elephants} by John, Kistler, which was the first book on the subject published since \textit{Histoire Militaire des Elephants} in 1843.}

In ancient warfare, elephants were pivotal, serving roles as diverse as logistical support to being the vanguard of an army, instilling terror and disarray among enemy ranks. Their impressive size and strength not only made them formidable shock troops but also provided a protective bulwark for infantry against enemy assaults. Unlike war horses, which primarily served as mounts, elephants possessed the intelligence to act independently, making decisions and taking actions with minimal human commands. This autonomy was leveraged in numerous historical campaigns, from the battlefields of Persia to the Roman Empire. Perhaps most famously, they were a hallmark of Hannibal's audacious crossing of the Alps into Italy. During the turbulent fourth and third centuries BC, following Alexander the Great's death, a veritable arms race ensued among his successors to build ever more powerful elephant forces. This period, known as the \enquote{Succession Wars,} saw an intense competition as each faction sought to tip the scales in their favor with the strength and psychological impact of elephants.

In the next sections, we will explore how to apply lessons from history to contemporary AI combat systems, illustrating the synergy between human teams of \enquote{AI mahouts} and battlefield AI. This examination will delve into adaptive strategies, social learning capabilities, and the handling of unpredictable battlefield scenarios. We will also assess defense against novel or 'zero day' exploits by adversaries and the psychological impact these AI-enabled systems might have on opposition forces. We will conclude with a discussion of the distinct roles and responsibilities assigned to AI operators and combatants in the field.

\subsection{AI + Humans are \textit{Adaptable}}
\label{sec:ai_and_humans_adapt}
Although they are quite capable of behaving autonomously, elephants are most effective when part of a human-in-the-loop (HITL) systems. The operator, or \textit{mahout}, operates the elephant much like a modern-day machine. Masters and their beasts interact through a complex system of commands, body language, and mutual understanding. To be part of an effective weapons system, an elephant needs a close relationship with its mahout.

\textit{Centaur} systems, have been a part of human/AI teaming since Garry Kasparov introduced the idea of \enquote{advanced chess} in 1998~\cite{kasparov2017deep}. These systems combine human input with machine learning algorithms to allow for more flexible and general solutions~\cite{scharre2016centaur}. The human provides insights and intuition that may be outside the context that the machine was designed and trained for. Creative, successful solutions from one operator can spread to others, greatly increasing the capability of the centaur beyond the sum of its parts~\cite{krakowski2022artificial}. Such systems provide a more adaptable solution to complex tasks that can be difficult or impossible for machines to solve on their own. The flexibility of the hybrid system makes it more robust, as it can quickly and easily adjust to changing conditions. 

\subsection{Social Learning}
\label{sec:social_learning}
In addition to individual instruction, it is crucial for teams comprised of elephants and their mahouts to operate in sync. Historically, as far back as 600 BC, troops that incorporated elephants cohabitated in designated settlements. This close proximity fostered a sense of community among the mahouts and their elephants, enhancing the efficiency with which these units could be mobilized for military purposes and facilitating mutual, social, learning.

This concept of social learning is not exclusive to humans and animals; it holds significant potential for reframing the field of artificial intelligence. It brings into relief how traditional methods of AI training, fundamental though they may be, diverge from the rich, nuanced learning processes seen in social creatures.

Social learning stands in contrast to the development process of current AI systems. Unlike artificial intelligence that is singularly developed and relies on extensive datasets for training—a process that can be both costly and time-consuming—humans (and elephants) benefit from the ability to absorb knowledge through societal interactions and cooperative efforts. This method of learning provides a level of flexibility and adaptability not present in isolated AI systems. When individuals with diverse experiences, backgrounds, and viewpoints come together, it enriches the learning environment. Such diversity spurs creativity and fosters innovation, offering a wellspring for AI applications that can cleverly address the intersecting challenges of technology and humanit.

\subsection{AI Unpredictability}
\label{sec:ai_unpredictability}

Like AI models, elephants are trained as well as possible for what their military expects. But when an elephant encounters something new, it may behave unpredictably. When Antongonus Gonatas besieged the city of Megara in 266 BC, the people of the city took pigs, doused them in sticky tar, set them on fire and had them run through the elephants. The elephants panicked and fled, which broke the siege. Thereafter Gonatas' elephants were housed with pigs among the elephants. 

Models can act only on what they have been trained on. A good model can generalize \textit{within} its training, but  it is unlikely to generalize \textit{outside} of it. In the case of the Megaran pigs, the real world can sometimes present unexpected challenges that require the AI system to adapt and adjust its behavior. An AI system trained to recognize and classify objects in photographs might be very accurate when it is tested on a set of curated images. However, if it is deployed to the real world and encounters photographs that are blurry, poorly lit or filled with background clutter, it will struggle. 

For combat AI systems, there is always an element of risk when using them in the field. They may not be able to predict or respond to certain situations that have not been accounted for in their training. It is important for designers of AI systems to consider not just the specific tasks the system is expected to complete, but also the wider environment in which the system will be deployed, and to allow for some level of flexibility and adaptability when designing the system.

\subsection{AI Swarm Behavior}
\label{sec:ai_swarms}
AI drone swarms have the potential to revolutionize various fields, including military operations. These systems rely on the coordinated behavior of a large number  vehicles which can work together to achieve a common goal.

However, the performance of an AI drone swarm may degrade in unpredictable ways as it loses communication or member drones. This is similar to the battle between Alexander and Darius in 331 BC, where King Darius' war elephants were surrounded by Alexander's forces and crowded into a narrow space where they couldn't keep formation and lost their mahouts. In this situation, the elephants lashed out in all directions, causing confusion and chaos in their own ranks that outweighed any damage done to Alexander's forces.

Similarly, if an AI drone swarm loses its communication or is forced into a difficult position, it may struggle to maintain its coordinated behavior and become disoriented. This disorientation could lead to erratic behavior, such as drones attacking each other or undesignated targets. To avoid these unintended consequences, it is important to ensure that any AI swarm maintains sufficient communication to coordinate and is not placed in environments outside its training domain.  Thus, it is critical to employ mechanisms to detect when the swarm is degrading and take appropriate action such as disengaging affected weapons systems.

\subsection{Zero-Day Exploits}
\label{sec:zero_day}

Mamood, the founder of the Ghaznavid Empire which spanned much of modern-day Iran, Afghanistan, and Pakistan was known for his use of war elephants in the Indian subcontinent. During Mamood's 1023 AD campaign against the Raja of Kalanger, saboteurs in the service of the Raja were able to slip intoxicating drugs into the War Elephants' water supply. The animals went on a rampage in the camp, causing \enquote{chaos and destruction}. Mamood ordered the elephants killed in order to prevent further damage, but the mahouts were able to gain enough control of the animals to lead them out into the forest to recover~\cite{ranking1826historical}.

While lethal autonomous weapons systems (LAWS) offer the potential for increased efficiency and precision on the battlefield, they also come with risks that are intrinsic to their autonomy. An enemy  able to manipulate a LAWS such as the Iranian Shahed drone or the US Switchblade loitering munition so that the weapons would turn and strike their own operators would, in addition to the psychological effects of such a scenario, create significant logistical impacts. Entire systems would have to be withdrawn from combat and examined for potential security exploits, leading to potential delays and disruptions in military operations.

\subsection{Expense}
\label{sec:expense}

There is a Nepalese Adage: \enquote{\textit{To take revenge on an enemy, buy him an elephant.' At first the recipient of your gift will be humbled by your generosity. Then he will be made a pauper as the elephant eats him out of house and home.}}

Training and updating models for autonomous weapons systems is a complex and time-consuming process that requires significant resources. For these systems to be effective, they must be able to accurately identify and classify targets, which requires large amounts of data and advanced algorithms. This process is made even more complex by the fact that the potential combat environment is constantly changing. This means that the models must be regularly updated in order to remain accurate and effective.

There is also the issue of fixing unforeseen problems that may arise. These systems are often highly complex, and it is not uncommon for such systems to encounter problems that are difficult to anticipate or resolve. This can be particularly problematic in the case of lethal autonomous weapons systems, as any errors or malfunctions could have serious consequences.

In addition, it is important to consider training data as its own consideration in the development of autonomous weapons systems. In order to train and test these systems, large amounts of data are required, which can be expensive and time-consuming to gather or to generate. The data will inevitably contain biases that may negatively affect the behavior of the weapons system, so it is important to ensure that the data is updated to be representative of any target environment and that models are continuously validated against real-world scenarios. Although history provides no clear guidance in the case of expensive models, at least it's good to know that this problem is not \textit{new}.

\subsection{Psychological Weapons}
\label{sec:psyops}

The psychological effects of weapons in warfare are multifaceted and often contribute to their overall efficacy. Flamethrowers, for instance, are a type of weapon whose psychological impact on both adversaries and operators is well documented~\cite{mcnab2015flamethrower}. Similarly, lethal autonomous weapons systems can be designed not only for physical destruction but also to inflict psychological strain on the enemy. 

One of the most powerful psychological effects of such systems is their ability to demoralize and intimidate enemy forces. This effect can be seen in the battle of King Purus versus Alexander at Hydaspes in 326 BC, in which Alexander's infantry had to face two hundred war elephants, an opponent that they had no experience with at the time. The elephants caused significant damage and broke the morale of Alexander's forces, who later threatened mutiny rather than face the beasts again.

In a modern context, this scenario echoes the potential effect of autonomous weapons on the battlefield. These systems' ability to act with unparalleled speed and precision could overwhelm human soldiers, planting the seed of futility in their minds. Much like Alexander's men who viewed the war elephants as an insurmountable threat, troops today might experience a parallel sense of demoralization when facing robotic adversaries perceived as unbeatable.

However, this element of surprise may work only against an unprepared adversary. In the battle of Gaza between Ptolemy and Seleucus against Demetrius in 312 BC, Demetrius used elephants as a screen for his cavalry.  But Seleucus and Ptolemy had encountered elephants before and their forces were not phased by their size and strength. Seleucus took his horses around Demetrius' elephants and was able to reach the unprotected enemy flank. He offered an honorable surrender to any who wished to quit the fight, and many divisions  of Demetrius' army accepted.


\subsection{Operators, not Warfighters}
\label{sec:operators_not_warfighters}
The Greek king Pyrrhus of Epirus in his battles against the Roman Empire is credited with introducing a new concept in the use of war elephants in battle: the separation of responsibilities between \textit{operating} the elephant by the mahout, and \textit{fighting} from the elephant with soldiers using weapons such as javelins, spears and arrows from protective towers on the backs of the elephants.

The combination of mahouts and protective towers was an innovation that greatly increased the combat capacity of Pyrrhus' war elephants. The shock and confusion caused by elephants charging into battle with warriors mounted on their backs was a formidable sight, and the Romans were reportedly terrified. The combination of mahouts and protective towers allowed the elephants to be used more effectively in combat, and their presence on the battlefield provided a tactical advantage that was difficult to overcome.

Similarly, when combat AI is used, it is important to divide responsibilities between the AI, the warfighter, and the operators. In this arrangement, In this case, the mahouts would be a mix of remote human operators augmented by AI operator proxies located in the battlespace, are teamed for the continuous, real-time monitoring of the battlefield AI at a neural network level. This enables them to detect any deviations or struggles the AI might have in following the established rules of engagement due to the unpredictable nature of warfare environments. 

The AI operator proxies work in tandem with, and learn from, human controllers, effectively \enquote{shadowing} their human counterparts to gain insights and understanding of nuanced decision-making processes. This unique setup ensures that, should traditional communication channels be disrupted -- whether through jamming or other methods -- the AI operator proxies are capable of independently managing the battlefield AI, thanks to their ongoing training and understanding gained from human operators.

Moreover, to safeguard against any potential misuse or erratic behavior by the AI, additional safeguards should be put in place. These would include sets of explicit guidelines that dictate the AI's actions under conditions that exceed the battlefield AI's ability to handle. For instance, a uniform anomaly detected across a significant portion of the AI models' neural behaviors should result in a command for the battlefield AI to immediately disengage from combat activities. This critical measure ensures that, even in the absence of direct human oversight, the AI's actions remain in strict compliance with international humanitarian laws.

By having operators monitor and oversee the battlefield AI, we can increase the likelihood that there is always an ethical component to autonomous combat scenarios. The operator would always ensure that the AI is behaving in line with the applicable laws. This division of responsibilities not only applies to scenarios where battlefield AI system mistakenly engages an invalid target, but scenarios where the AI would fail to engage appropriately given the context. This operator interaction with the system frees the warfighter to focus on tasks such as mission planning, strategy, and tactical execution. This approach would allow military organizations to take advantage of the benefits of AI while still maintaining a human element in highly dynamic, often chaotic processes.

\subsection{Additional lessons}
As befits a history that covers thousands of years, there are many lessons to be learned from the study of war elephants. Elephants fight each other in ways that are unlike fighting humans or horses. This implies that when autonomous engage each other, that distinctive patterns will emerge. We will need to be prepared for that so that models and their humans can adjust. 


Another item -- mahouts would often get their animals drunk before combat. This made them more aggressive but also harder to control. We should consider having every combat AI contain multiple models that can be switched on or off depending on the context. There may be many times where a model developed to avoiding contact with the enemy may be preferable to one that engages. In very special conditions, it may be desirable to engage models with unrestricted targeting. All these high-level behavior modifications need to be under direct human control and should probably have innate timeout mechanisms.

	\section{DISCUSSION}
\label{sec:discussion}


In 2022, artificial intelligence reached an inflection point. The conflict in Ukraine forced a rapid evolution of autonomous drones as electronic warfare intensified.  Meanwhile, the release of powerful language models like ChatGPT captivated a global audience, even as their limitations became starkly apparent. These systems, prone to factual errors and outright fabrications, underscored the risk in mistaking pattern recognition for genuine understanding. As AI disrupts entire industries, we're learning that its fluency often masks a fundamental inability to distinguish truth from fiction.\footnote{\url{https://storage.courtlistener.com/recap/gov.uscourts.nysd.575368/gov.uscourts.nysd.575368.32.1_1.pdf}}

The military use of AI presents unique and thorny ethical questions. AI systems, trained on specific data, are powerful tools for their designed purpose. They excel at pattern recognition and the rapid execution of learned tasks. What remains less clear is an AI system's ability to adapt to novel situations, especially those where human lives are at stake.

\begin{figure}[!htbp]
    \centering
    \fbox{\includegraphics[width=18em]{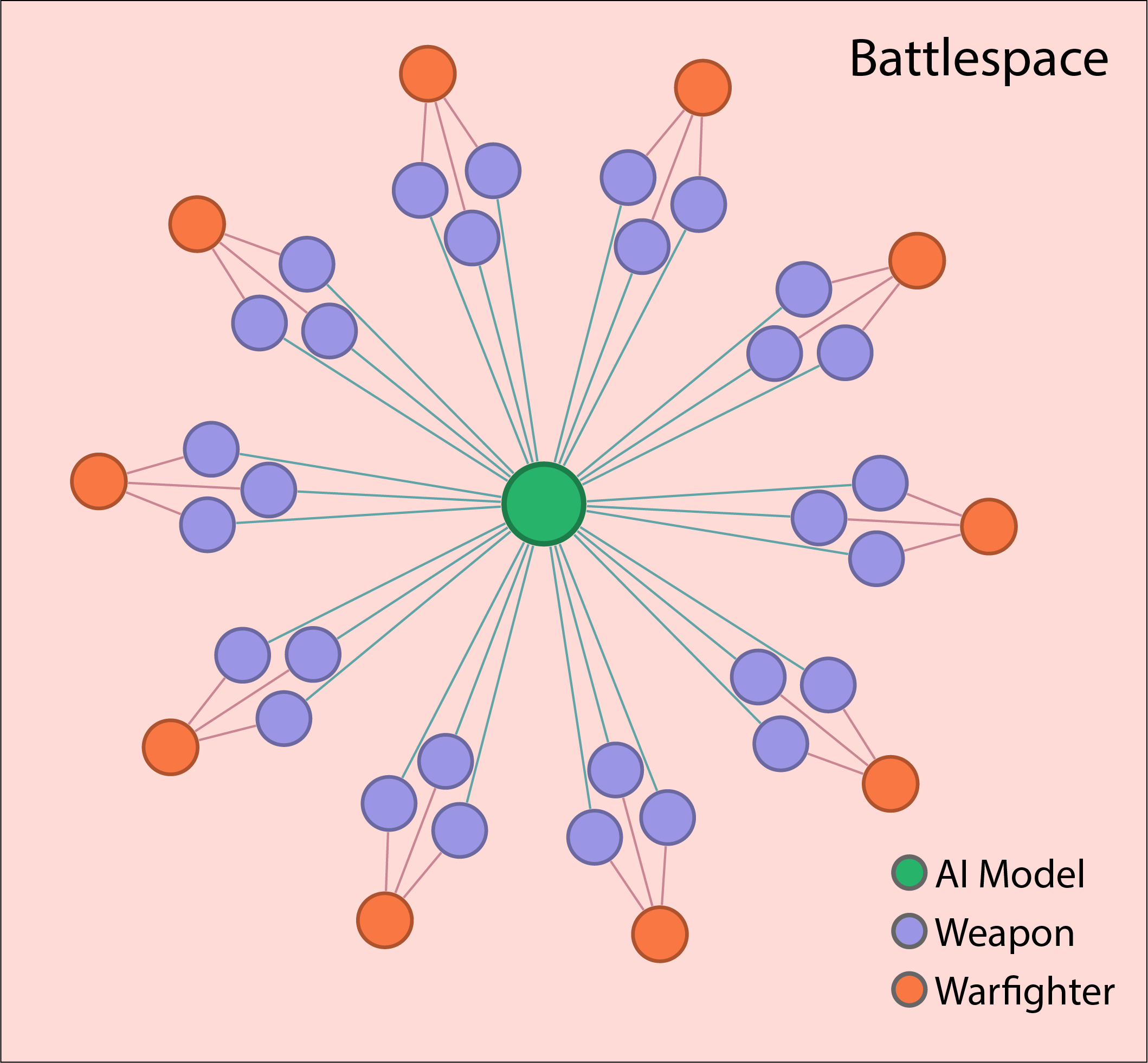}}
    \caption{\label{fig:brittle_combat_ai}Brittle Combat AI}
\end{figure}

While AI offers undeniable power in data processing and analysis, its ability to guide behavior within an ethical framework is less certain. Consider Figure~\ref{fig:brittle_combat_ai}, that shows a canonical organization of autonomous battlefield AI. A warfighter (orange circles) directly operates a suite of weapons systems (blue circles), each mediated through a single AI model, trained to control the particular weapon to engage the enemy. However, if the enemy has worked out ways to fool the AI model, the warfighter often cannot change the behavior of the model. Rather the warfighter must fall back on other, potentially riskier means to engage the enemy.  The emphasis on simplified, streamlined AI models within warfare introduces a dangerous brittleness. When an AI model is fooled, it could place warfighters at heightened risk while simultaneously ceding a critical advantage to the enemy.

In this paper, we suggest the role of an AI operator or \enquote{mahout} as introduced in section~\ref{sec:operators_not_warfighters}. This role extends the concept of the \enquote{centaur} introduced by Garry Kasparov in the context of \enquote{advanced chess,} where humans team up with multiple chess programs, blending human strategic insight with machine computational power to compete at extraordinarily high levels~\cite{scharre2016centaur, kasparov2017deep}. These areas of mutual reinforcement, or \textit{complementation} have been shown to outperform either one of the components alone~\cite{tang2020chiron}. The idea of complementation, where multiple elements are combined to produce a whole that is more effective than any one of its parts, is fundamentally different from the more common practice of \textit{substitution}, where the most capable component -- human, automation, or AI -- is used in isolation. The practical outcome of substitution systems is that they are only as strong as the weakest component. 

\begin{figure}[!htbp]
    \centering
    \fbox{\includegraphics[width=22em]{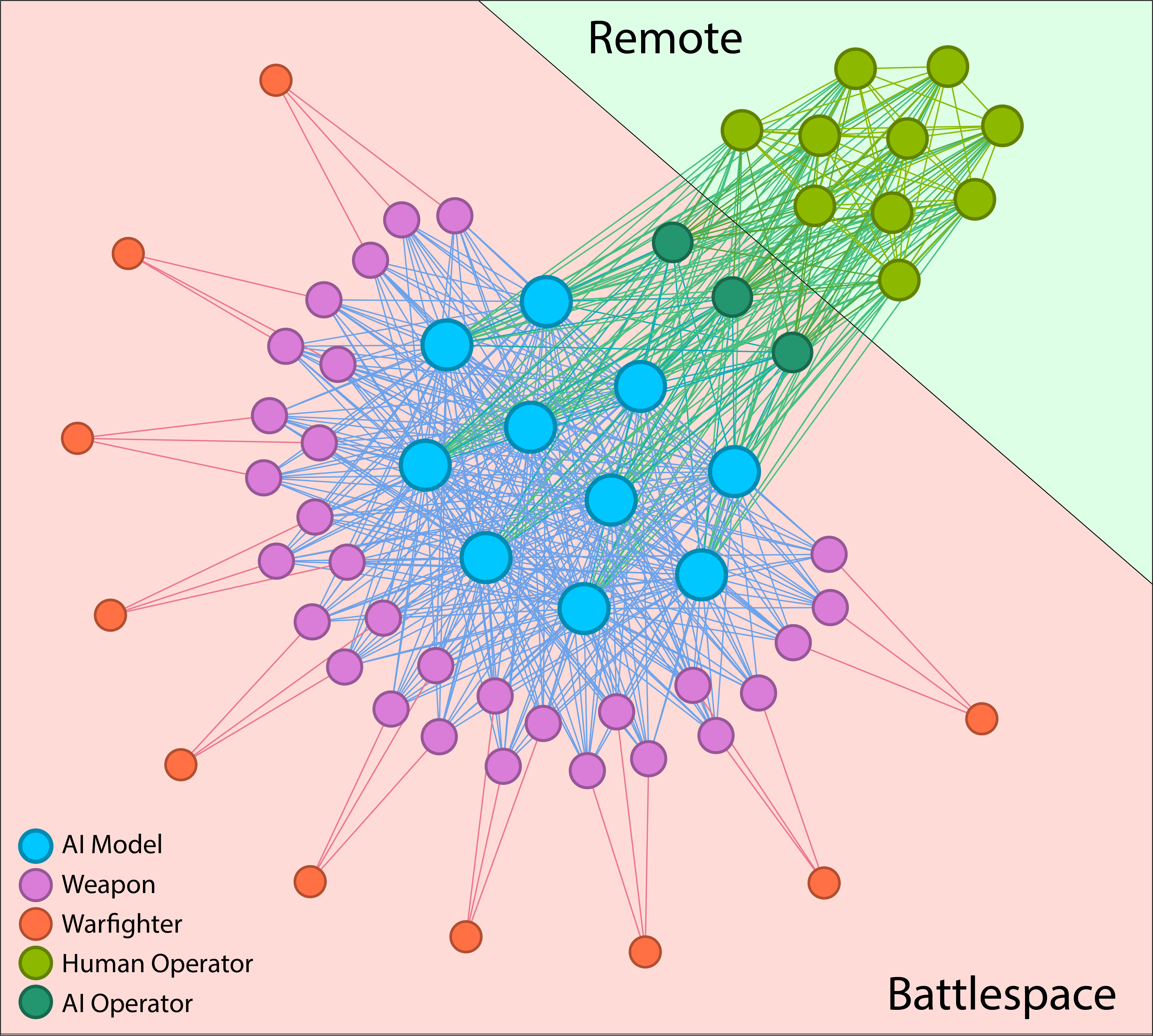}}
    \caption{\label{fig:resiliant_combat_ai}Resilient Combat AI}
\end{figure}

The network of warfighter, battlefield AI, remote human operator, and their local AI proxies ensure that the strengths of both humans and AI are leveraged to their fullest (Figure~\ref{fig:resiliant_combat_ai}). This evolution of the centaur concept into military strategy illustrates a future where human-AI collaboration is not just beneficial but essential for achieving superiority in highly dynamic and competitive environments. 

For AI systems to function reliably in military contexts, human operators must understand the AI's decision-making processes. This means training operators with diverse skills and backgrounds. Like chess players adjusting their strategy, operators should be able to switch AI models as needed. A single AI model cannot be expected to function perfectly across every type of battlefield scenario. Conditions are unpredictable; the best model for one situation may be ineffective in another. Operators must understand the strengths and weaknesses of each AI model to ensure they are using the most effective tool for the task at hand. This adaptability is critical to the successful integration of AI into military operations.

Adopting this flexible approach to pairing humans with AI in military operations creates systems inherently more nimble and adaptable than those operated by AI alone. This human-AI symbiosis can disrupt an AI-only adversary's Observe, Orient, Decide, Act (OODA) loop~\cite{osinga2007science}, effectively \enquote{turning inside} it. AI, enriched by human intuition and experience, introduces unpredictability and complexity that a purely AI adversary might struggle to comprehend or process in real time. This could decisively tilt the balance in favor of human-AI teams, providing them with a tactical edge by outmaneuvering and outthinking AI-only systems in critical moments.

Humans can use their creativity and innovation to come up with new ideas and solutions that AI can't. For example, incorporating human knowledge of known hazards and risks into autonomous systems allows operators to determine and predict the behavior of the AI-based controller when encountering real-world risk factors~\cite{DREANY20181}. We can also understand the nuances and context of a situation that AI misses, which can be critical in decision-making. Lastly, humans perform best in groups, where each person brings different perspectives and skills, providing a \textit{depth of understanding} that current AI appears to mimic, but cannot replicate.

\subsection{Switchable models}

Given the vulnerabilities outlined in sections \textit{\nameref{sec:zero_day}} and \textit{\nameref{sec:ai_unpredictability}}, the potential for AI models to behave unexpectedly in novel situations is a major concern. To mitigate this risk, human/AI-proxy operators should be able to monitor model performance closely and identify potential failures. It's also crucial to have a diverse set of models on hand, allowing for flexibility in unforeseen circumstances.  Where necessary, operators must be able to disable AI functionality altogether or switch to deterministic or rule-based systems. Ideally, multiple AI models would run concurrently, providing operators with comparative data to inform their decisions during rapidly evolving scenarios

The human/AI-proxy operator holds a complementary position to AI systems. Like a mahout, the operator's role is to ensure the AI stays focused on its intended task, not to fight. This lets the operator focus on managing multiple models in real-time, scrutinizing their outputs for anomalies or potential errors, and selecting the most appropriate model given the situation. Should problematic behavior arise, the operator would be empowered to override or disable the malfunctioning AI.

By handling the technical aspects of the AI system, the operator frees the warfighter or commander to make decisions informed by trustworthy, vetted, machine-generated insights. Successful integration of AI into military operations depends on this complementary dynamic, maximizing both human judgment and the rapid analytical power of machines.

\subsection{Decoding the AI Battlefield}
\label{sec:human_solutions}

We are no longer the only intelligence in the battlespace. The rise of artificial intelligence (AI) means we share the battlespace with a new kind of entity. Understanding this evolving landscape, and how AI systems perceive us, may define the future of military operations. To navigate this complexity, humans will require a powerful mix of analytics, visualization tools, and AI itself. These technologies enable real-time analysis and decision-making, providing an indispensable edge in the rapid flow of combat.

However, success hinges on the seamless connection between these technological capabilities and the missions they support.  A mismatch between a system's design and the task at hand can lead to misinterpretations and costly errors. Similarly, overloading personnel with too much responsibility in this high-stakes environment can jeopardize the entire operation. Finding the right balance between human expertise and machine augmentation is crucial.

\begin{figure}[!htbp]
    \centerline{
	\centering
    \begin{subfigure}{.5\textwidth}
    	\centering
    	\fbox{\includegraphics[height=10em]{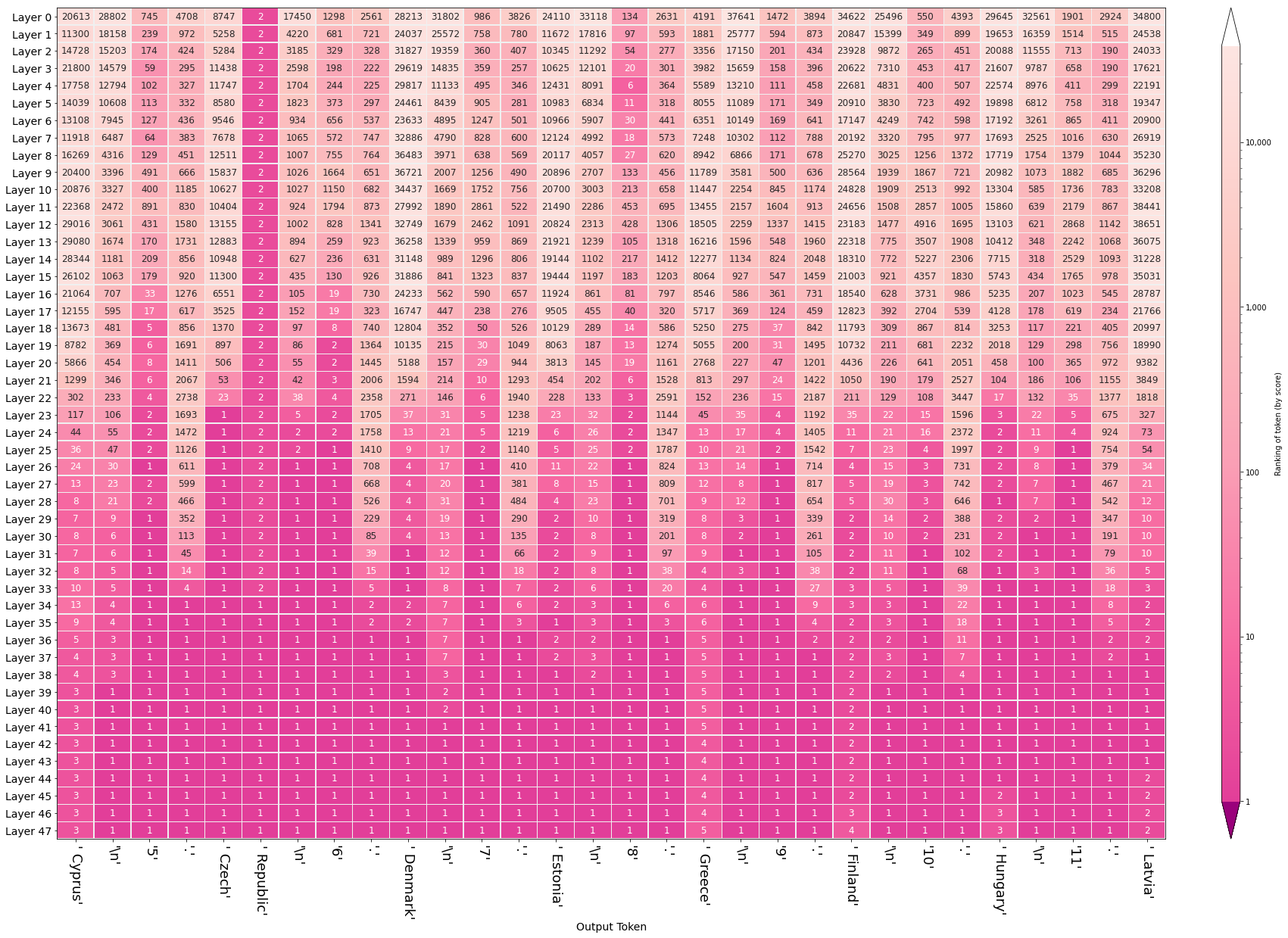}}
    	\caption{\label{fig:GPT-countries}GPT Layer Activations~\cite{alammar2021hiddenstates}}
    \end{subfigure}
    \begin{subfigure}{.5\textwidth}
    	\centering
    	\fbox{\includegraphics[height=10em]{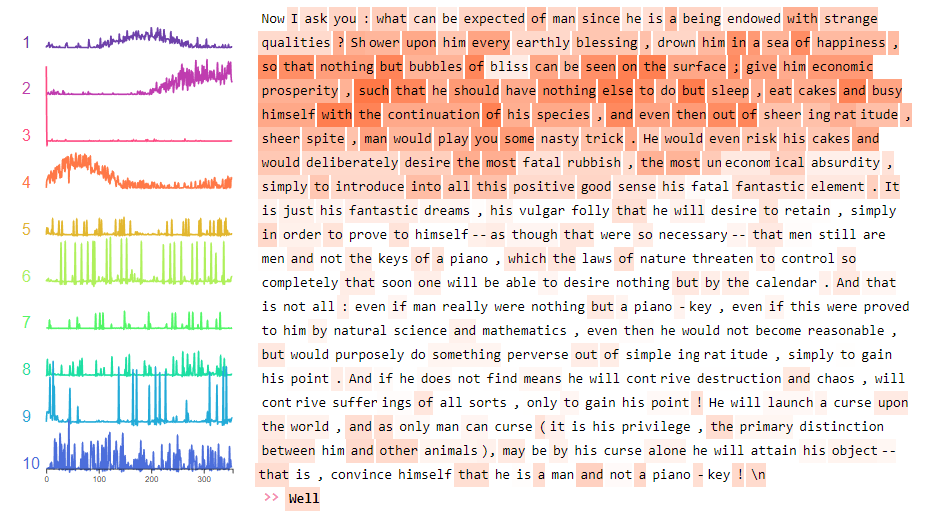}}
    	\caption{\label{fig:GPT-factors}GPT Layer Factors~\cite{alammar2020explaining}}
     \end{subfigure} 
    }
    \vspace{1em}
    \centerline{
     \begin{subfigure}{.75\textwidth}
    	\centering
    	\fbox{\includegraphics[height=12em]{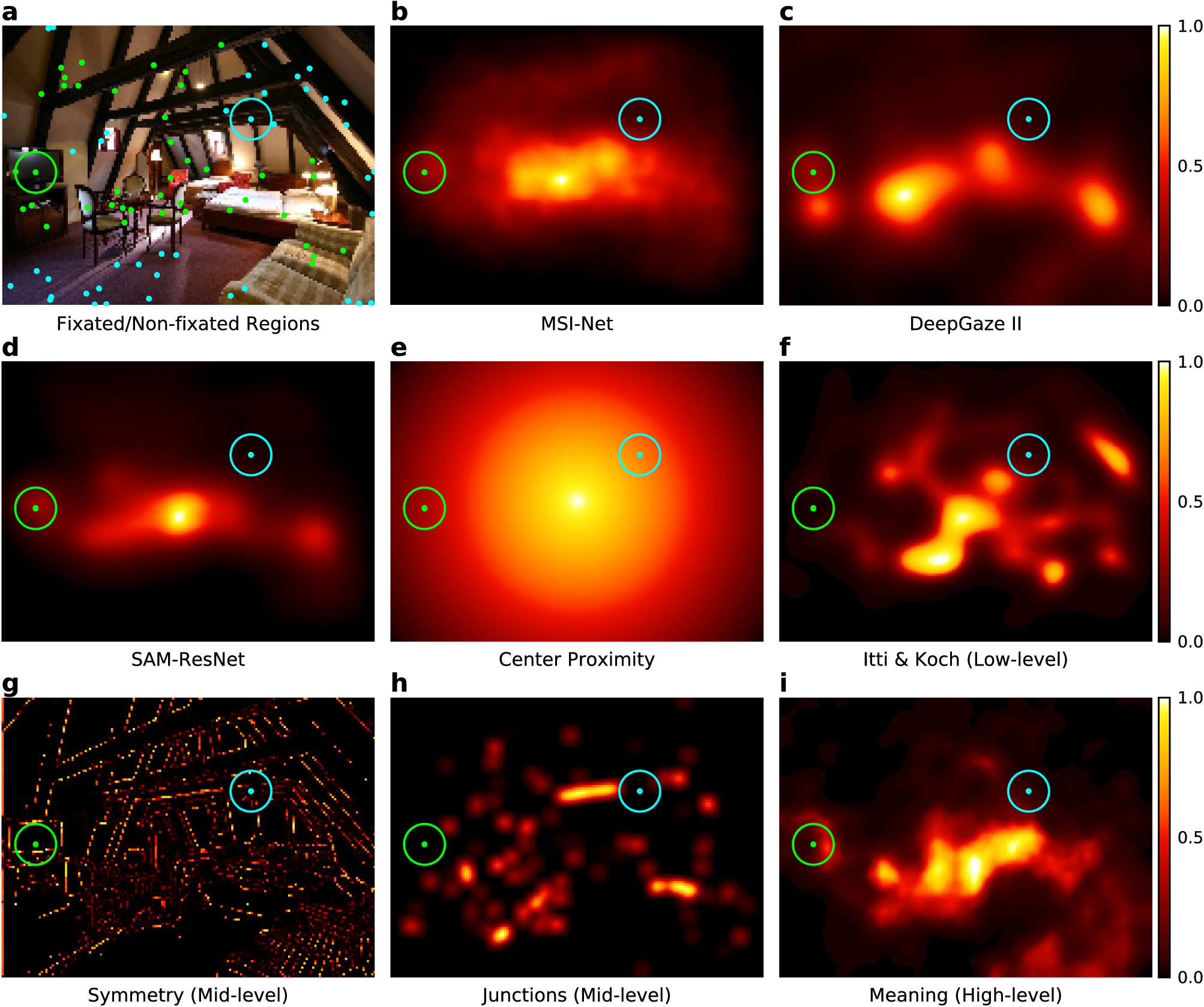}}
    	\caption{\label{fig:salience_map}Deep Salience Maps~\cite{hayes2021deep}}
     \end{subfigure}
    }
    \caption{Interfaces for Model Behavior}
    \label{fig:model_interfaces}
\end{figure}

To understand when we should trust our increasingly sophisticated machines, operators need visualizations tailored to understanding the domain of the layers, weights, and activation thresholds of these  systems. For the best results, visualizations must also be tailored to the individual user’s domain knowledge and understanding of the system. When the goal becomes \textit{to determine if the AI working}, then salience and factor displays like Figures~\ref{fig:GPT-countries}, \ref{fig:GPT-factors}, and \ref{fig:salience_map} may be far more important than the more traditional displays.

Such displays can show in real time how a model is reacting. A good model will behave in ways that are complex -- not overly simple and not chaotic. Multiple models trained to perform the same tasks may behave in different ways, and one may clearly be best. It should be possible for the operators to seamlessly swap models to always keep the most effective model active. Input from trained warfighters and commanders should be incorporated into the system as well -- they may not be able to perform at the speed and scale of the actual systems, but could provide valuable insight into the overall behavior. The goal is to provide tools that can be used to verify that the system is behaving as intended and to take corrective action if not. 

Providing the necessary insight to allow humans to understand the system and its behavior will require a combination of analytics, visualization, and AI. The goal is to enable humans to make better-informed decisions in a timely manner. Such real-time decision support requires the development of visualization tools that enable users to quickly assess the performance and trustworthiness of AI-driven systems. 

Modern warfare is becoming increasingly complex and requires new understanding of technology, analytics, and AI. It is essential that humans remain in control of these complex systems. Developing visualization tools that enable humans to understand the system’s behavior and take corrective action when necessary is critical for maintaining trust and ensuring success in untested environments.

\subsection{Mechanical Mahouts}
\label{sec:mechanical_mahouts}

Understanding the role of humans in an environment of ubiquitous AI is vital to understand the future of LAWS. While the idea of lethal systems operating without human control rightfully evokes unease, there will be times when communication with battlefield AI will not be possible. The key rests not in blind faith in machines, but in a human-machine partnership -- a framework where humans guide AI, and AI fills in the gaps when human intervention becomes impossible. By understanding that AI excels at pattern-based responses, the way is opened to developing reliable supervisory AI companions for LAWS. However, this concept hinges on a critical partnership: the human Operator. Human intelligence -- particularly group intelligence -- remains unmatched in creativity, adaptation, and social learning.

An Operator AI system, modeled on the concept of \enquote{shadow driving} in the self-driving car industry, trains in tandem with its human partners. Through simulations and live training, the Operator AI absorbs the patterns, tactics, and strategies employed by those it serves. When combat necessitates autonomous operation, this Operator AI  becomes a 'mechanical mahout' -- guiding the LAWS to continue the mission within the framework built during human-supervised periods.

This understanding of AI's capabilities is the cornerstone for developing LAWS that can operate ethically in dynamic combat scenarios. The key lies in leveraging AI's strength in pattern mimicry to support battlefield supervision and decision-making, particularly when traditional human control is untenable due to electronic warfare or other forms of disruption. By crafting AI systems that can maintain the pattern of human decision-making in combat scenarios, it becomes possible to construct a semblance of human oversight, even when direct control is not feasible.

However, this approach necessitates a sophisticated architecture that seamlessly integrates AI's capabilities with human judgement. Such systems would essentially function on two levels. On one level, the Operator AI would operate in a passive role, monitoring the actions of the human Operators and potentially alerting them when it detects behavior of the battlefield AI that seems misaligned with the rules of engagement. On the other, the Operator AI steps in when the remote human Operators cannot maintain supervisory control, maintaining the patterns that have developed over the engagement and previous training. This dual-level approach ensures that, despite operating autonomously, LAWS remain tethered to a human decision-making framework, embodying the 'human in the loop' ethos even when direct human control is not possible.

Even so, as much as these models excel in pattern recognition, they can drift from initial tasks, which could lead to decisions that diverge from established  guidelines or rules of engagement. This potential divergence poses a tangible risk to the ethical application of lethal force. A failsafe is essential, one that uses a more deterministic, rules-based system. Conventional software avoids the stochastic wanderings of neural networks, and can be depended on to adhere strictly to predefined rules of engagement. 

This failsafe would be engaged based on the  AI's behavior and operational duration without human supervision. Predetermined criteria, based on the system's performance and the length of unsupervised operation, would trigger this shift. For example, in the face of ambiguous or conflicting information, the failsafe may require the withdrawal of the LAWS from the battlespace. This recall isn't an admission of failure but a testament to the commitment to ethical warfare. It signifies the understanding that when AI drifts beyond the realm of human oversight, the system must be realigned with human values and rules of engagement. 

The goal of integrating AI into lethal autonomous weapons systems must be to augment human decision-making, not replace it. The melding of human oversight with AI's capabilities offers the potential for enhanced operational efficiency and effectiveness. By nesting AI within a framework of human oversight -- indirect though it may be -- we can create a framework for systems that are not only autonomous but also adhere to guidelines and rules of engagement. This approach addresses the critical challenge of maintaining ethical accountability in autonomous systems. Such a framework acknowledges AI's limitations while maximizing its capabilities.

	\section{CONCLUSIONS}

\begin{displayquote}
\textit{\textbf{Recommendation: Develop innovative human-centric approaches to human-machine teaming}. The kind of data fusion envisioned here through autonomous machine-to-machine integration will require new concepts for human-machine teaming that optimize the strengths of each}~\cite{schmidt2021national}.

\hfill -- National Security Commission on Artificial Intelligence (Final)
\end{displayquote}

Integrating AI into combat systems can significantly impact warfare and the role of autonomous weapons. System trust, where humans rely too heavily on machines, is a concern for military AI systems. Excessive trust can lead to poor decision-making. To address this human limitation, AI combat systems should incorporate diverse groups of people to operate the AI component using AI-centric displays, while warfighters supervise the weapons using appropriate interfaces. This approach aligns with the NSCAI final report.

Using human/AI teams mirrors the use of War Elephants, where complementation was key to achieving a greater result than either alone. A Mahout controlled the elephant, while soldiers fought from protective towers on the animals' backs. Similarly, a human-in-the-loop approach based on complementation can ensure resilient and effective use of combat AI. Humans contribute originality, social problem solving, and contextualization, while AI offers precision, consistency, and speed~\cite{krakowski2022artificial}.

There are fundamental reasons that the concept of complementation, where machines and humans work together in a tight loop is more capable and resilient than an approach relying on substitution, where a human sits on the loop, waiting to take over if the AI behaves incorrectly. To illustrate the advantages of complementation over substitution, consider AI/human interaction as a network~\cite{lewis2022many}. In Figure~\ref{fig:brittle_combat_ai} in Section~\ref{sec:discussion}, there is a single AI model that is trained for use in  the deployed weapon. One warfighter simultaneously operates three weapons as a human-on-the-loop increasing the liklihood of cognitive overload during dynamic combat situations. With only one AI model that is replicated over all the deployed devices, an adversary need only find a single exploit to render the entire system ineffective.

Alternatively, Figure~\ref{fig:resiliant_combat_ai} in Section~\ref{sec:discussion} shows an example of the more resilient approach described in this paper. Here, a set of AI models, each with a different architecture and/or training data is selected by a diverse group of Operators. Weapons are supervised, rather than controlled by warfighters. Rather than a single point of failure, the Operators, models, and weapons are connected by a dense network. The Operators are responsible for ensuring that the AI models being used by the weapons are performing correctly. If they are not, another can be swapped in. The complementary inclusion of human beings makes the system capable of adapting to situations not considered in the model training. Multiple options reduce the attack surface by making it much more difficult for an adversary to disable all the weapons by focusing on a single aspect of the system. 

In both cases, the warfighter's core mission remains the control of their LAWS. What has fundamentally shifted is their role within a complex interplay of human judgment and machine intelligence.  The transition from technology substituting for the warfighter to complementing them reduces the risk of cognitive overload. Warfighters now oversee instead of micromanaging. Their decisions ripple through a network of AI models, each offering distinct capabilities. This bolsters confidence in the system's resilience to failure and its ability to adapt to the unpredictable nature of combat. 


Automated systems are limited to their configuration and training, which cannot be updated on the fly. Integrating people with varying backgrounds and perspectives opens up dimensions that are unique to each person. Rather than trying to turn inside another machine's observe, orient, decide, act (OODA) loop, the addition of complementary human Operators opens up new pathways and opportunities for success on the battlefield. More importantly, this approach allows for the AI/human team to be continuously updated and improved in real-time, as it is not limited to predefined configurations that no longer match a changing environment.

The concept of complementation has a strong ethical and moral component, as it requires the AI to be designed and operated in a way that allows for the human to maintain control and oversight. In contrast, approaches where the AI is allowed to make decisions largely in its own may result in ethical or legal violations. As such, the use of complementation is not only technically beneficial, but also ethically defensible, as it ensures that humans retain control of the system on multiple levels and can be held accountable for the outcomes of their decisions.


As technology expanded the battlefield to include cyber security, the need for \enquote{Cyber Warriors} arose. Similarly, the modern battlefield is expanding with the introduction of AI weapons~\cite{fulp2003training}, requiring the creation of an AI Operator role, or a modern Mahout, to train and equip them. Figure~\ref{fig:model_interfaces} shows examples of tools used to monitor AI behavior and determine the appropriate model for use in different contexts. The AI Operator would ensure responsible and intended use of the AI, requiring a deep understanding of its workings and responsible decision-making.


Human Operators and their AI proxies are essential for deploying and operating AI systems as a way of ensuring appropriate levels of human judgement on the behavior of LAWS, particularly in conditions of limited communication. Operators monitor the behavior of battlefield LAWS, troubleshoot and resolve issues, and ensure they function as intended during exercises and combat. With the increasing complexity of AI systems, there is a higher risk of failure in unforeseen conditions. Even well-trained models may exhibit unexpected or undesirable behavior in such situations. A diverse team of AI Operators can bring unique insights to finding solutions.

Combat AI stands apart from the economics-driven AI common in industry. In the commercial realm, the focus is on streamlining processes and maximizing profits -- goals best served by a single, optimized model that meets the broadest possible customer need. But in warfare, where circumstances are unpredictable, a single AI solution could easily fail or be misled. One of the key responsibilities of Operators would be to recognize when a model is struggling and to find a better alternative in an ensemble of models that were trained using different data sets or architectures, or models that are better suited to handling specific types of data or conditions. 



The Operator bears the accountability for evaluating the effectiveness and proportionality of the AI models in use. If an appropriate model cannot be identified, it becomes the responsibility of the Operator or their proxy to deactivate the battlefield AI until it can be restored to nominal behavior or replaced with a deterministic system. This measure is necessary to prevent unintended or disproportionate harm caused by the model. The Human Operators serve as crucial advisors in the development and testing of new models. Given their familiarity with the potential weaknesses and limitations of these models, they provide valuable insights and recommendations for enhancing performance.

The role of Operators will be crucial in ensuring that AI models are used ethically and effectively in combat. By monitoring the behavior of these models and being prepared to take corrective action when necessary, Operators would ensure that models would always perform in compliance with IHL, and would be able to deliver on their full potential while providing real value to the warfighter.

It is important to recognize the potential benefits and limitations of AI in combat systems and to carefully consider the ethical implications of its use. A human-centered approach, in which the strengths of both humans and AI are balanced, is likely to be the most effective way forward for the foreseeable future. Operators play a key role in this approach, ensuring that LAWS are always used in accordance with the principles of international humanitarian law.

	\section*{Disclaimer}
	The views expressed in this paper are those of the authors and do not reflect the official policy or position of the US Navy, Department of Defense or the US Government.
	
	\newpage

\end{document}